\documentclass[aps,pra,groupedaddress,floats,reprint]{revtex4-2}

\usepackage{amsfonts,graphicx,soul,natbib,mathrsfs,dsfont,amssymb}

\usepackage{amsmath}
\usepackage{graphicx}
\usepackage{amsthm}
\usepackage{array}
\usepackage{physics}
\usepackage{makecell}
\usepackage[dvipsnames]{xcolor}
\usepackage{colortbl}
\usepackage{mathtools}
\usepackage{stmaryrd}
\usepackage{tcolorbox}
\usepackage{bm}
\usepackage[colorlinks=true, allcolors=blue]{hyperref}

\begin{document}

\title{Fringe field induced spin-OAM mixing of twisted electrons}

\author{S. V. Gatalina}%
\affiliation{Institute of Physics and Mechanics, Peter the Great St. Petersburg Polytechnic University, Polytechnicheskaya 29, Saint Petersburg 195251, Russia}%

\author{S. S. Baturin}%
\affiliation{School of Physics and Engineering,
ITMO University, St. Petersburg 197101, Russia}%

\date{\today}

\begin{abstract} We study spin effects in twisted-electron propagation through the entrance or exit region of an axially symmetric magnetic coil. Starting from the Foldy-Wouthuysen reduction of the Dirac equation, we derive a paraxial spinor equation in which the longitudinally varying solenoidal field produces, in addition to the usual diagonal Zeeman term, a transverse Pauli coupling proportional to the fringe-field gradient. The scalar transverse dynamics is treated exactly by the Ermakov mapping, which absorbs the longitudinally dependent focusing into a metaplectic scaling transformation and reduces the orbital evolution to that of a stationary two-dimensional oscillator. On this background, the transverse Pauli term is treated perturbatively and yields an explicit first-order correction for arbitrary realistic solenoidal profiles. Axial symmetry implies conservation of the total projection of angular momentum, so each spin flip is accompanied by a compensating one-quantum change of orbital angular momentum. In addition, the linear coordinate structure of the perturbation restricts the first-order dynamics to at most two neighboring radial sidebands for each incoming oscillator component. We derive the corresponding transition amplitudes and show how their phases are governed jointly by the Ermakov accumulation and the diagonal spin-orbital rotation. The resulting framework provides a direct way to quantify spin-orbit mixing of twisted electrons in realistic magnetic lenses and solenoidal beam-line elements, and it identifies a route toward controlled spin-OAM conversion in engineered sequences of magnetic-field edges. \end{abstract}


\maketitle

\section{Introduction}

Electron vortex beams are free-electron states with a well-defined
azimuthal phase winding and nonzero intrinsic orbital angular momentum
(OAM).  Since their theoretical proposal and first experimental
realizations, they have become an important platform for studying the
interplay between wave-front topology, magnetic moments, and structured
electron-matter interactions
\cite{Bliokh2007,Uchida2010,Verbeeck2010,McMorran2011}.  Large-OAM
electron beams have been generated in transmission electron microscopes,
including implementations reaching more than one thousand quanta of
orbital angular momentum.  These developments have motivated applications
in magnetic and chiral contrast, electron energy-loss spectroscopy,
nanoscale magnetic mapping, and the manipulation of matter by structured
electron beams \cite{Lloyd2017,Bliokh2017}.

At the same time, twisted charged-particle states are of growing interest
beyond the microscopy regime, including scattering and radiation processes
at higher energies, where the orbital-angular-momentum degree of freedom
can modify angular-momentum selection rules and observable distributions
\cite{Ivanov2011,Ivanov2022}. This has motivated efforts aimed at
extending vortex-particle physics to accelerator-compatible energies and
geometries, as well as dedicated studies of radiation and scattering
observables for structured electrons and other vortex particles
\cite{Karlovets2023,Chaikovskaia2024,Pavlov2025,Sheremet2025}. It has
also stimulated the development of compact source concepts for vortex
electrons, including recent injector-oriented proposals
\cite{Liu2025,Dyatlov2025,RFPhotoinjector2025}. From this
perspective, understanding the propagation of twisted electrons through
realistic magnetic lenses and solenoidal elements is a necessary step
toward controlled generation, transport, and manipulation of such states
in beam-line environments.

Most theoretical descriptions of twisted-electron propagation through
magnetic lenses focus on the scalar orbital dynamics \cite{VanEnk2021,Filina2024,Filina2026}.  This is often an
adequate leading approximation.  In a uniform longitudinal magnetic field,
the Pauli term is diagonal in the spin basis and produces the usual
Zeeman phase \cite{PhysRevA.103.L010201}.  Real magnetic lenses and solenoids, however, are not
uniform.  Their entrance and exit regions necessarily contain transverse
magnetic-field components fixed by the longitudinal field gradient \cite{Filina2026}.  The
corresponding transverse Pauli interaction couples spin to the transverse
coordinates.  For a vortex state, the transverse Pauli term acts directly on the azimuthal
structure of the wave function and induces spin-orbit transitions.

The purpose of this work is to formulate this spin-fringe effect
quantitatively for twisted electrons propagating through an axially
symmetric, longitudinally inhomogeneous solenoidal field.  Starting from
the Foldy-Wouthuysen reduction  \cite{FW,itzykson1985quantum,SilenkoFW,SilenkoG}
of the Dirac equation, we derive the
paraxial spinor propagation equation including the full near-axis Pauli
coupling.  The scalar part of the problem is treated exactly by the
Ermakov mapping \cite{Aldaya_2011,PhysRevA.108.012219,Filina2026}, which absorbs the longitudinally dependent focusing into
a metaplectic scaling transformation.  The transverse Pauli term is then
treated perturbatively on top of this exact scalar evolution.

The axial symmetry of the fringe field plays a central role.  The
spin-fringe perturbation does not conserve the orbital and spin
projections separately.  Instead, it conserves their sum, the total axial
angular momentum.  As a result, a spin flip must be accompanied by a
compensating one-quantum change of the orbital angular momentum.  For each
initial oscillator component, the radial structure of the linear
coordinate perturbation further restricts the transition to at most two
neighboring oscillator-shell channels.

The resulting first-order correction is obtained as an explicit
propagation integral over the experimentally specified solenoidal field
profile.  Its magnitude is controlled by the fringe-gradient strength
dressed by the Ermakov scaling factor, while its phase contains both the
scalar Ermakov accumulation and the diagonal spin-orbital rotation.  This
form is applicable to arbitrary realistic solenoidal profiles satisfying
the paraxial and perturbative conditions.  It therefore provides a direct
way to estimate spin-orbit mixing in magnetic lenses of electron
microscopes and in solenoidal elements used in accelerator beam lines.

The present analysis also connects to the broader question of intrinsic
multipole properties of twisted electrons.  It has been argued that
vortex spin-\(1/2\) particles can possess a measurable electric quadrupole
moment and tensor magnetic polarizability \cite{Silenko2019}.  The
spin-fringe transitions derived here provide a concrete dynamical channel
through which the structured transverse profile of a twisted electron
becomes coupled to spin in a realistic magnetic-lens geometry.  This makes
solenoidal fringe fields a natural setting for probing spin-orbit effects
beyond the scalar theory of vortex-electron transport.

\section{Paraxial model in a solenoidal fringe field}

We consider a charged Dirac particle propagating along the axis of an
axially symmetric magnetic lens.  In the near-axis region the magnetic field is determined, to first order in the
transverse displacement, by the on-axis profile \(B_z(z)\):
\begin{align}
\label{eq:near_axis_B}
\mathbf B(x,y,z)
&=
\left(
-\frac{x}{2}\frac{dB_z}{dz},
-\frac{y}{2}\frac{dB_z}{dz},
B_z(z)
\right)
+\mathcal{O}(\rho^2), \\ \nonumber
\rho^2&=x^2+y^2 .
\end{align}
Thus the entrance and exit regions of the coil necessarily contain a
transverse magnetic field proportional to \(dB_z/dz\).  It is precisely
this fringe component which produces the spin-coordinate coupling studied
below.

We use the regular cylindrically symmetric Coulomb-gauge representative
\begin{equation}
\mathbf A(x,y,z)
=
\frac{B_z(z)}{2}(-y,x,0),
\label{eq:A_coulomb}
\end{equation}
for which \(\nabla\cdot\mathbf A=0\), \(A_z=0\), and
\(\nabla\times\mathbf A\) reproduces Eq.~\eqref{eq:near_axis_B} to the
order retained in the near-axis expansion.

For a stationary state of energy \(E\), the squared Dirac equation, or
equivalently the Foldy-Wouthuysen reduction \cite{FW,itzykson1985quantum,SilenkoFW,SilenkoG} for the upper spinor
component, gives the Pauli-type equation
\begin{equation}
\left[
\hat{\boldsymbol\pi}^{\,2}
-
q\,\bm\sigma\cdot\mathbf B
\right]\Phi
=
k^2\Phi,
\label{eq:pauli_squared}
\end{equation}
with 
\begin{align}
k^2=E^2-m^2,~~
\hat{\boldsymbol\pi}=\hat{\mathbf p}-q\mathbf A 
\end{align}

where \(q\) is the particle charge and \(\hbar=c=1\).  

We introduce a substitution
\begin{align}
    \Phi(\mathbf r)=e^{ikz}\varphi(\mathbf r)
\end{align}

and assume that the paraxial condition is fulfilled 
\begin{align}
    |\partial_z^2\varphi|\ll k|\partial_z\varphi|,~~
p_\perp^2\ll k^2 .
\end{align}

Then Eq.~\eqref{eq:pauli_squared} reduces to the Schr\"{o}dinger-type equation for $\varphi$
\begin{equation}
i\frac{\partial\varphi}{\partial z}
=
\frac{1}{2k}
\left[
\hat{\boldsymbol\pi}_\perp^{\,2}
-
q\,\bm\sigma\cdot\mathbf B
\right]\varphi .
\label{eq:physical_paraxial}
\end{equation}

The orbital part of the kinetic term is
\begin{align}
\hat{\boldsymbol\pi}_\perp^{\,2}
&=
\hat p_\perp^2
-
qB_z(z)\hat L_z
+
\frac{q^2B_z^2(z)}{4}\hat \rho^2, \nonumber\\ 
\hat L_z&=\hat{x}\hat p_y-\hat{y}\hat p_x .
\label{eq:pi_expand}
\end{align}
The Pauli term naturally separates into a longitudinal Zeeman part and a
fringe-induced transverse part
\begin{equation}
-q\,\bm\sigma\cdot\mathbf B
=
-qB_z(z)\sigma_z
+
\frac{q}{2}\frac{dB_z}{dz}
\left(
\hat{x}\sigma_x+\hat{y}\sigma_y
\right).
\label{eq:pauli_split}
\end{equation}
Equivalently,
\begin{equation}
\hat{x}\sigma_x+\hat{y}\sigma_y
=
\hat{\rho} e^{-i\sigma_z\phi}\sigma_x,
\label{eq:spin_orbit_fringe_operator}
\end{equation}
Here $\phi$ is the polar angle.

We now introduce dimensionless transverse variables and a dimensionless
propagation coordinate,
\begin{equation}
\tilde{x}=\frac{x}{\rho_H},~~
\tilde{y}=\frac{y}{\rho_H},~~
\tilde{z}=\frac{z}{k\rho_H^2},
\label{eq:dimensionless_variables}
\end{equation}
with
\begin{align}
\rho_H=\sqrt{\frac{2}{|q|B_0}},~~
B_0=\max|B_z(z)|,
\end{align}
and define
\begin{equation}
\Omega(z)=\frac{B_z(z)}{B_0},
\qquad
s_q=\operatorname{sgn}(q).
\label{eq:omega_def}
\end{equation}

To keep the notation clean we now switch to the normalized coordinates and drop tildes.

After dropping tildes, the paraxial equation becomes
\begin{align}
\label{eq:dimensionless_spinful}
&i\frac{\partial\varphi}{\partial z}
= \\ \nonumber
&\left[
\hat{\mathcal H}_\perp(z)
-
s_q\Omega(z)\hat L_z
-
s_q\Omega(z)\sigma_z
+
\hat W(z)
\right]\varphi ,
\end{align}
where
\begin{align}
\label{eq:Hperp_def}
\hat{\mathcal H}_\perp(z)
=
\frac{\hat p_\perp^2}{2}
+
\frac{\Omega^2(z)\hat \rho^2}{2},
\end{align}
and the spin-fringe term is
\begin{align}
\hat W(z)
= 
\frac{s_q}{2k\rho_H}\Omega'(z)\,
\hat \rho e^{-i\sigma_z\phi}\sigma_x .
\label{eq:W_def}
\end{align}
Here and below the prime denotes differentiation with respect to the
dimensionless coordinate \(z\).

The transverse Pauli term Eq.\eqref{eq:W_def} commutes with the total angular momentum
\[
\hat J_z=\hat L_z+\hat s_z.
\]
Above 
\begin{equation}
\hat s_z=\frac{\sigma_z}{2},
\end{equation}
where $\sigma_z$ is a Pauli $z$ matrix.
Indeed, Eq.\eqref{eq:W_def} once rewritten as 
\begin{align}
\label{eq:Wzpl}
\hat W(z)
=
\frac{s_q}{2k\rho_H}\Omega'(z)\hat\rho
\left(
e^{-i\phi}\sigma_+
+
e^{i\phi}\sigma_-
\right)
\end{align}
satisfies
\begin{align}
\label{eq:comW}
[\hat W(z),\hat J_z]=0.
\end{align}
Above we used standard notation for the Pauli ladder matrices
\begin{align}
    \sigma_\pm=\frac{\sigma_x\pm i\sigma_y}{2}.
\end{align}
As follows from Eq.\eqref{eq:Wzpl} and Eq.\eqref{eq:comW} the fringe field does not break cylindrical symmetry.  Rather, it
couples the two basis states within a fixed \(J_z\) sector,
\[
|l,\downarrow\rangle \leftrightarrow |l-1,\uparrow\rangle,
\qquad
|l,\uparrow\rangle \leftrightarrow |l+1,\downarrow\rangle .
\]
The spin-fringe term therefore mixes spin and orbital degrees of freedom
while preserving the conserved total projection \(m_j=l+s\).

The diagonal orbital and spin rotations commute with the scalar
oscillator Hamiltonian,
\begin{equation}
[\hat{\mathcal H}_\perp(z'),\hat L_z]=0,
~~
[\hat{\mathcal H}_\perp(z'),\sigma_z]=0,
~~
\forall z',
\label{eq:commuting_terms}
\end{equation}
whereas the Hamiltonians \(\hat{\mathcal H}_\perp(z')\) and
\(\hat{\mathcal H}_\perp(z'')\) do not commute for a nonuniform field.
We therefore remove diagonal rotation first
\begin{align}
\label{eq:rotation_transform}
&\varphi(z)
=
\exp\left[
is_q\Theta(z)(\hat L_z+\sigma_z)
\right]\psi(z), \\ \nonumber
&\Theta(z)=\int_0^z\Omega(z')\,dz' .
\end{align}
The transformed spinor obeys
\begin{align}
\label{eq:rotated_equation}
i\frac{\partial\psi}{\partial z}
=
\left[
\hat{\mathcal H}_\perp(z)
+
\hat V_R(z)
\right]\psi,
\end{align}
with
\begin{align}
\label{eq:VR_def}
\hat V_R(z)
=
&\frac{s_q}{2k\rho_H}\Omega'(z)\hat\rho \times \\ \nonumber
&\left[
e^{-i(\phi+s_q\Theta(z))}\sigma_+
+
e^{i(\phi+s_q\Theta(z))}\sigma_-
\right].
\end{align}
We note that the phase $s_q\Theta(z)$ in Eq.~\eqref{eq:VR_def} is a rotating-frame phase. The transverse Pauli term
commutes with the physical axial generator 
\(\hat J_z=\hat L_z+\hat s_z\), while the diagonal part removed in
Eq.~\eqref{eq:rotation_transform} is generated by
\(\hat L_z+2\hat s_z\).  The mismatch between these two generators leaves
the residual phase multiplying the spin-flip operator.

Equations \eqref{eq:rotated_equation} and \eqref{eq:VR_def} along with Eq.\eqref{eq:rotation_transform} constitute the main model that we study further. 

We note that dimensionless parameter $1/(k \rho_H)$ is naturally small \cite{EpovArxiv}. Indeed, for typical transmission-electron-microscope parameters, $B_0 \sim 1$ T and $W_e \sim 100$ keV we have for the characteristic magnetic length $\rho_H=36.3$ nm and for the modulus of the electron wave vector $k=1.7$ pm$^{-1}$. This results in $1/(k \rho_H)\approx 10^{-5}$. Therefore we may treat the operator $\hat V_R(z)$ as a parametrically suppressed perturbation to the Hamiltonian $\hat{\mathcal H}_\perp(z)$. 

\section{Ermakov mapping}

We briefly recall the Ermakov mapping \cite{Aldaya_2011,PhysRevA.108.012219,Filina2026} 
generated by
\begin{equation}
i\frac{\partial\psi}{\partial z}
=
\hat{\mathcal H}_{\perp}(z)\psi,
\qquad
\hat{\mathcal H}_{\perp}(z)
=
\frac{\hat p_\perp^2}{2}
+
\frac{\Omega^2(z)\hat\rho^2}{2}.
\label{eq:ermakov_start}
\end{equation}
Although this Hamiltonian is quadratic, it is explicitly \(z\)-dependent
at the entrance and exit regions of the solenoid.  The Ermakov
transformation maps Eq.~\eqref{eq:ermakov_start} to a stationary
two-dimensional oscillator,
\begin{equation}
i\frac{\partial\psi_0}{\partial \tau}
=
\hat H_0\psi_0,
\qquad
\hat H_0
=
\frac{\hat p_{0\perp}^2}{2}
+
\frac{\hat\rho_0^2}{2}.
\label{eq:H0_def}
\end{equation}
The scaled transverse coordinate and the intrinsic propagation parameter
are
\begin{equation}
\rho_0=\frac{\rho}{b(z)},
\qquad
\tau(z)=\int_0^z\frac{d\zeta}{b^2(\zeta)}.
\label{eq:ermakov_variables}
\end{equation}
The corresponding wave functions are related by
\begin{align}
\label{eq:ermakov_wavefunction}
&\psi(\rho,\phi,z)
= \\ \nonumber
&\frac{1}{b(z)}
\exp\left[
i\frac{b'(z)}{2b(z)}\rho^2
\right]
\psi_0
\left(
\frac{\rho}{b(z)},\phi,\tau(z)
\right).
\end{align}
Here the factor \(1/b(z)\) is the two-dimensional normalization factor,
while the quadratic phase represents the wave-front curvature induced by
the varying focusing strength.

The scaling function \(b(z)\) satisfies the Ermakov-Pinney equation
\begin{equation}
b''(z)+\Omega^2(z)b(z)
=
\frac{1}{b^3(z)},
\label{eq:ermakov_pinney}
\end{equation}
with initial conditions
\begin{equation}
b(0)=b_0,
\qquad
b'(0)=b'_0 .
\end{equation}
In the normalization introduced in Eq.\eqref{eq:dimensionless_variables} and Eq.\eqref{eq:omega_def}, the maximum-field plateau of the solenoid has \(\Omega=1\).  Therefore a beam matched to this uniform region
corresponds to
\begin{equation}
b(z_u)=1,
\qquad
b'(z_u)=0,
\end{equation}
where $z_u$ is taken away from the fringes inside the solenoid.

The Ermakov mapping may be written in operator form as well
\begin{equation}
\psi(z)=\hat{\varepsilon}(z)\psi_0(\tau(z)),
\label{eq:operator_mapping}
\end{equation}
where
\begin{align}
\label{eq:ermakov_operator}
&\hat{\varepsilon}(z)
= \\ \nonumber
&\exp\left[
i\frac{b'(z)}{2b(z)}\hat\rho^2
\right]
\exp\left[
-\frac{i}{2}\ln b(z)
\left(
\hat{\boldsymbol\rho}\cdot\hat{\mathbf p}_\perp
+
\hat{\mathbf p}_\perp\cdot\hat{\boldsymbol\rho}
\right)
\right].
\end{align}
Above $\hat{\boldsymbol\rho}$ is the transverse position vector.

This operator implements the coordinate scaling and the quadratic
curvature phase in Eq.~\eqref{eq:ermakov_wavefunction}.  The stationary
Hamiltonian is obtained from
\begin{equation}
\hat H_0
=
b^2(z)
\left[
\hat{\varepsilon}^{-1}(z)
\hat{\mathcal H}_{\perp}(z)
\hat{\varepsilon}(z)
-
i\hat{\varepsilon}^{-1}(z)
\frac{\partial\hat{\varepsilon}(z)}{\partial z}
\right].
\label{eq:H0_from_epsilon}
\end{equation}

Thus the Ermakov mapping absorbs the scalar breathing dynamics into the
metaplectic operator \(\hat{\varepsilon}(z)\) and leaves a stationary
unit-frequency oscillator as the reference problem.  

\section{Perturbation theory}

After the diagonal orbital and Zeeman rotations have been removed, the
spinor \(\psi\) satisfies
\begin{equation}
i\frac{\partial\psi}{\partial z}
=
\left[
\hat{\mathcal H}_\perp(z)+\hat V_R(z)
\right]\psi ,
\label{eq:pert_start}
\end{equation}
where $\hat{\mathcal H}_\perp(z)$ is given by Eq.\eqref{eq:Hperp_def} and the rotated spin-fringe perturbation $\hat V_R(z)$ by Eq.\eqref{eq:VR_def}

We now apply the Ermakov transformation with $\tau(z)$ given by Eq.\eqref{eq:ermakov_variables}
\begin{equation}
\psi(z)=\hat\varepsilon(z)\chi(\tau).
\end{equation}
The unperturbed scalar Hamiltonian $\hat{\mathcal H}_\perp(z)$ is mapped to the stationary oscillator $\hat H_0$ given by Eq.\eqref{eq:H0_def} and the transformed Eq.\eqref{eq:pert_start} becomes
\begin{equation}
i\frac{\partial\chi}{\partial\tau}
=
\left[
\hat H_0+\tilde V(\tau)
\right]\chi ,
\label{eq:tau_pert_eq}
\end{equation}
where
\begin{equation}
\tilde V(\tau)
=
b^2(z)
\hat\varepsilon^{-1}(z)
\hat V_R(z)
\hat\varepsilon(z)
\bigg|_{z=z(\tau)} .
\label{eq:Vtilde_def}
\end{equation}

Since the Ermakov operator rescales the transverse coordinate according to
\begin{equation}
\hat\varepsilon^{-1}(z)\hat\rho\hat\varepsilon(z)
=
b(z)\hat\rho_0,
\end{equation}
while leaving \(\phi\) and the spin matrices unchanged, the transformed perturbation 
takes the form
\begin{align}
\label{eq:Vtilde_explicit_tau}
\tilde V(\tau)
=
&\frac{s_q}{2k\rho_H}
b(z)\frac{d\Omega}{d\tau}\hat\rho_0\times \nonumber\\
&\left[
e^{-i(\phi+s_q\Theta(z))}\sigma_+
+
e^{i(\phi+s_q\Theta(z))}\sigma_-
\right]_{z=z(\tau)},
\end{align}
where we have used the identity
\begin{align}
\frac{d\Omega}{d\tau}=b^2(z)\Omega'(z).
\end{align}

We now pass to the interaction picture \cite{LandauQM,Sakurai2017} with respect to \(\hat H_0\)
\begin{equation}
\chi_I(\tau)=e^{i\hat H_0\tau}\chi(\tau).
\end{equation}
Then
\begin{equation}
i\frac{\partial\chi_I}{\partial\tau}
=
\hat V_I(\tau)\chi_I(\tau),
\end{equation}
where
\begin{equation}
\hat V_I(\tau)
=
e^{i\hat H_0\tau}
\tilde V(\tau)
e^{-i\hat H_0\tau}.
\end{equation}
Expanding
\begin{equation}
\chi_I(\tau)=\chi_I^{(0)}+\chi_I^{(1)}(\tau)+\cdots
\end{equation}
and imposing \(\chi_I^{(1)}(0)=0\), one obtains
\begin{equation}
\chi_I^{(1)}(\tau)
=
-i\int_0^\tau
\hat V_I(\tau')\chi_I^{(0)}\,d\tau' .
\label{eq:first_order_interaction}
\end{equation}
Returning to the Schr\"odinger representation of the Ermakov-mapped
problem gives
\begin{equation}
\chi^{(1)}(\tau)
=
-i\int_0^\tau
e^{-i\hat H_0(\tau-\tau')}
\tilde V(\tau')
\chi^{(0)}(\tau')\,d\tau' .
\label{eq:first_order_schrodinger}
\end{equation}
Finally, the first-order correction in the rotated physical frame is
obtained by applying the Ermakov operator:
\begin{equation}
\psi^{(1)}(z)
=
\hat\varepsilon(z)\chi^{(1)}(\tau(z)).
\label{eq:first_order_physical}
\end{equation}
The correction to the original paraxial spinor is
then recovered from
\begin{equation}
\label{eq:corrfin}
\varphi^{(1)}(z)
=
\exp\left[
is_q\Theta(z)(\hat L_z+\sigma_z)
\right]\psi^{(1)}(z).
\end{equation}
Although the unperturbed Ermakov dynamics is naturally expressed in the
intrinsic variable \(\tau\), the fringe profile is specified as a function
of the physical propagation coordinate \(z\).  We therefore keep \(z\) as
the integration variable and absorb the Jacobian \(d\tau/dz=1/b^2(z)\)
into the interaction-picture perturbation. Thus we adopt the following notation
\begin{align}
\label{eq:Vtilde_explicit_taumod}
\tilde V^{(z)}(z)
=
&\frac{s_q}{2k\rho_H}
b(z)\Omega'(z)\hat\rho_0\times \nonumber\\
&\left[
e^{-i(\phi+s_q\Theta(z))}\sigma_+
+
e^{i(\phi+s_q\Theta(z))}\sigma_-
\right].
\end{align}
The correction to the wave function in the interaction picture is then
\begin{equation}
\label{eq:first_order_z_final}
\chi_I^{(1)}(z)
=
-i\int_0^{z}
\hat V_I^{(z)}(\tilde{z})\,
\chi_I^{(0)}\,d \tilde{z} .
\end{equation}
In the Schr\"{o}dinger picture we have correspondingly
\begin{equation}
\label{eq:first_order_z_schrodinger}
\chi^{(1)}(z)
=
-i\int\limits_0^{z}
e^{-i\hat H_0[\tau(z)-\tau(\tilde{z})]}
\hat V^{(z)}(\tilde{z})
\chi^{(0)}(\tilde{z})
d\tilde{z}.
\end{equation}
For convenience, we have introduced the shorthand
\begin{align}
&\hat V^{(z)}(z)\equiv \tilde V^{(z)}(z), \nonumber
\\
&\hat V_I^{(z)}(z)
=
e^{i\hat H_0\tau(z)}
\hat V^{(z)}(z)
e^{-i\hat H_0\tau(z)}.
\end{align}
Equation \eqref{eq:first_order_z_schrodinger} with Eq.\eqref{eq:first_order_physical} and Eq.\eqref{eq:corrfin} allows one to calculate the correction to the wave function of the original problem \eqref{eq:dimensionless_spinful} in real coordinates, when operator $\hat{W}$ is parametrically suppressed by the smallness of $1/(k \rho_H)$.  

\section{Correction to the wave function}

In the Ermakov-mapped variables, the unperturbed reference Hamiltonian is just an oscillator Hamiltonian in Eq.\eqref{eq:H0_def}. Its normalized eigenfunctions in polar coordinates are
\begin{equation}
\psi^c_{n,l,s}(\rho_0,\phi)
=
N_{n,l}
\rho_0^{|l|}
L_n^{|l|}(\rho_0^2)
e^{-\rho_0^2/2}
e^{il\phi}
g_s,
\end{equation}
where superscript $c$ stands for cylindrical coordinates,  
\begin{equation}
g_+
=
\begin{pmatrix}1\\0\end{pmatrix},
\qquad
g_-
=
\begin{pmatrix}0\\1\end{pmatrix},
\end{equation}
and the normalization factor is given by
\begin{equation}
N_{n,l}
=
\sqrt{
\frac{n!}{\pi(n+|l|)!}
}.
\end{equation}
The corresponding eigenvalues are
\begin{equation}
\varkappa_{n,l}
=
2n+|l|+1 .
\end{equation}
They are independent of the spin projection, since the diagonal Zeeman
rotation has already been removed.

The general unperturbed state is therefore
\begin{align}
\label{eq:zero_ord}
\chi^{(0)}(\rho_0,\phi;z)
=
\sum_{n,l,s}
c_{n,l,s}
\psi^c_{n,l,s}(\rho_0,\phi)
e^{-i\varkappa_{n,l}\tau(z)},
\end{align}
where $\tau(z)$ is the Ermakov ``time" given by Eq.\eqref{eq:ermakov_variables}.

\begin{widetext}

Since the fringe profile is naturally given as a function of the
propagation coordinate \(z\), we write the first-order correction as a
\(z\)-integral.  The perturbation entering this integral is given by Eq.\eqref{eq:Vtilde_explicit_taumod} and we plug the zeroth-order wave function Eq.\eqref{eq:zero_ord} in to the Eq.\eqref{eq:first_order_z_schrodinger} for the correction in the Schr\"odinger representation.
\begin{align}
\label{eq:chifin}
\chi^{(1)}(\rho_0,\phi,z)
=
-\frac{i s_q}{2k\rho_H}
\sum_{n,l,s}
\sum_{n',l'}
c_{n,l,s}\,
\psi^c_{n',l',-s}(\rho_0,\phi)\,
e^{-i\varkappa_{n',l'}\tau(z)}
\Lambda^{(s)}_{n'l';nl}
\mathcal I^{(s)}_{n'l';nl}(z),
\end{align}
where
\begin{equation}
\mathcal I^{(s)}_{n'l';nl}(z)
=
\int_0^{z}
b(\tilde{z})\Omega'(\tilde{z})
\exp\left[
i\left(\varkappa_{n',l'}-\varkappa_{n,l}\right)\tau(\tilde{z})
+
2is\,s_q\Theta(\tilde{z})
\right]d\tilde{z} .
\end{equation}
\end{widetext}
For \(l\geq1\), the nonzero matrix elements are
\begin{equation}
\Lambda^{(-1/2)}_{n'l';nl}
=
\delta_{l',l-1}
\left[
\sqrt{n+l}\,\delta_{n',n}
-
\sqrt{n+1}\,\delta_{n',n+1}
\right],
\label{eq:negspl}
\end{equation}
corresponding to
\[
|n,l,\downarrow\rangle
\rightarrow
|n',l-1,\uparrow\rangle,
\]
and
\begin{equation}
\Lambda^{(+1/2)}_{n'l';nl}
=
\delta_{l',l+1}
\left[
\sqrt{n+l+1}\,\delta_{n',n}
-
\sqrt{n}\,\delta_{n',n-1}
\right],
\end{equation}
corresponding to
\[
|n,l,\uparrow\rangle
\rightarrow
|n',l+1,\downarrow\rangle .
\]
For \(l\leq -1\), the nonzero matrix elements are
\begin{equation}
\Lambda^{(-1/2)}_{n'l';nl}
=
\delta_{l',l-1}
\left[
\sqrt{n+|l|+1}\,\delta_{n',n}
-
\sqrt{n}\,\delta_{n',n-1}
\right],
\label{eq:negsl}
\end{equation}
corresponding to
\[
|n,l,\downarrow\rangle
\rightarrow
|n',l-1,\uparrow\rangle,
\]
and
\begin{equation}
\Lambda^{(+1/2)}_{n'l';nl}
=
\delta_{l',l+1}
\left[
\sqrt{n+|l|}\,\delta_{n',n}
-
\sqrt{n+1}\,\delta_{n',n+1}
\right],
\end{equation}
corresponding to
\[
|n,l,\uparrow\rangle
\rightarrow
|n',l+1,\downarrow\rangle .
\]
For \(l=0\), the two spin-flip channels give
\begin{equation}
\Lambda^{(-1/2)}_{n'l';n0}
=
\delta_{l',-1}
\left[
\sqrt{n+1}\,\delta_{n',n}
-
\sqrt{n}\,\delta_{n',n-1}
\right],
\end{equation}
corresponding to
\[
|n,0,\downarrow\rangle
\rightarrow
|n',-1,\uparrow\rangle,
\]
and
\begin{equation}
\Lambda^{(+1/2)}_{n'l';n0}
=
\delta_{l',1}
\left[
\sqrt{n+1}\,\delta_{n',n}
-
\sqrt{n}\,\delta_{n',n-1}
\right],
\end{equation}
corresponding to
\[
|n,0,\uparrow\rangle
\rightarrow
|n',1,\downarrow\rangle .
\]
All spin-preserving matrix elements vanish.

Thus the spin-fringe perturbation obeys the selection rule
\begin{equation}
s'=-s,
\qquad
l'=l+2s,
\end{equation}
or equivalently
\begin{equation}
\Delta l+\Delta s_z=0.
\end{equation}
The total axial projection \(m_j=l+s\) is therefore conserved as expected.

Although \(\hat H_0\) has degenerate eigenspaces, the perturbation is
linear in the transverse coordinate and changes the oscillator shell by
one quantum.  Hence \(\Delta\varkappa=\pm1\), and no matrix element acts
within a degenerate eigenspace of \(\hat H_0\).  Degenerate perturbation
theory is therefore not required at first order.

The correction obtained above is written in the Ermakov-mapped
representation.  To recover the correction to the rotated paraxial spinor
$\psi$, we use the connection Eq.\eqref{eq:first_order_physical} and apply Ermakov operator Eq.\eqref{eq:ermakov_wavefunction}, Eq.\eqref{eq:ermakov_operator}.
This immediately gives
\begin{equation}
\psi^{(1)}(\rho,\phi,z)
=
\frac{1}{b(z)}
\exp\left[
i\frac{b'(z)}{2b(z)}\rho^2
\right]
\chi^{(1)}
\left(
\frac{\rho}{b(z)},\phi,z
\right).
\label{eq:psi1_from_chi1}
\end{equation}

\begin{widetext}

In the last step the original paraxial spinor $\varphi$ is recovered with the help of the Eq.\eqref{eq:corrfin} by switching back to the non-rotating frame.

Substituting Eq.\eqref{eq:psi1_from_chi1} into Eq.\eqref{eq:corrfin} and the explicit expansion for $\chi^{(1)}$ given by Eq.\eqref{eq:chifin} one obtains
\begin{align}
\varphi^{(1)}(\rho,\phi,z)
=
-\frac{i s_q}{2k\rho_H}
\frac{1}{b(z)}
\exp\left[
i\frac{b'(z)}{2b(z)}\rho^2
\right]
\sum_{n,l,s}
\sum_{n',l'}
&
c_{n,l,s}\,
\Lambda^{(s)}_{n'l';nl}\,
\mathcal I^{(s)}_{n'l';nl}(z)
\nonumber\\
\times\,
&
e^{-i\varkappa_{n',l'}\tau(z)}
e^{is_q\Theta(z)(l'-2s)}
\psi^c_{n',l',-s}
\left(
\frac{\rho}{b(z)},\phi
\right).
\label{eq:varphi1_compact}
\end{align}

Using the selection rule \(l'=l+2s\), the final rotation phase may be
simplified:
\begin{equation}
e^{is_q\Theta(z)(l'-2s)}
=
e^{is_q l\Theta(z)} .
\end{equation}
Thus an equivalent form where the selection rules are accounted for is
\begin{align}
\varphi^{(1)}(\rho,\phi,z)
=
-\frac{i s_q}{2k\rho_H}
\frac{1}{b(z)}
\exp\left[
i\frac{b'(z)}{2b(z)}\rho^2
\right]
\sum_{n,l,s}
\sum_{n'}
&
c_{n,l,s}\,
\Lambda^{(s)}_{n',\,l+2s;\,n l}\,
\mathcal I^{(s)}_{n',\,l+2s;\,n l}(z)
\nonumber\\
\times\,
&
e^{-i\varkappa_{n',\,l+2s}\tau(z)}
e^{is_q l\Theta(z)}
\psi^c_{n',\,l+2s,\,-s}
\left(
\frac{\rho}{b(z)},\phi
\right).
\label{eq:varphi1_selection_reduced}
\end{align}

\end{widetext}

\section{Discussion}

The first-order result derived above describes the formation of the
spin-fringe admixture during passage through a single solenoidal edge.
It can also be used as the starting point for treating several edges or a
longitudinally structured solenoidal profile.  The appropriate
generalization is not stationary perturbation theory, but the
\(z\)-ordered evolution generated by the interaction-picture perturbation
\(\hat V_I^{(z)}(z)\).

In the Ermakov interaction picture, the exact evolution operator is
\begin{equation}
\hat U_I(z_f,z_i)
=
\mathcal Z
\exp\left[
-i\int_{z_i}^{z_f}\hat V_I^{(z)}(z)\,dz
\right],
\label{eq:z_ordered_evolution}
\end{equation}
where \(\mathcal Z\) denotes ordering along the propagation coordinate.
Expanding this exponential gives the Dyson series
\begin{align}
&\hat U_I(z_f,z_i)
=
1
-i\int_{z_i}^{z_f}dz_1\,\hat V_I^{(z)}(z_1)
\\ \nonumber
&+(-i)^2
\int_{z_i<z_1<z_2<z_f}
dz_2\,dz_1\,
\hat V_I^{(z)}(z_2)\hat V_I^{(z)}(z_1)
+\cdots .
\label{eq:dyson_series_z}
\end{align}
The first term beyond unity reproduces the first-order correction obtained
above.  Higher-order terms describe repeated spin-fringe transitions,
either within one extended inhomogeneous region or through several
separated coil edges.  In the latter case, the propagation phases between
edges are automatically included by the \(z\)-ordered product.

A key simplification follows from axial symmetry.  Each insertion of
\(\hat V_I^{(z)}(z)\) commutes with the total axial angular momentum
\(\hat J_z=\hat L_z+\hat s_z\).  Consequently,
\begin{equation}
[\hat U_I(z_f,z_i),\hat J_z]=0 .
\end{equation}
Thus all orders of perturbation theory preserve the initial value of
\(J_z\).  Starting from a pure component \(|n,l,s\rangle\), the evolution
can populate only states satisfying
\begin{equation}
l'+s'=l+s .
\end{equation}
Since \(s'=\pm1/2\), a fixed \(J_z\) sector contains only two possible
orbital projections $l'=l$ for the original spin projection, and $l'=l+2s$ for the spin-flipped projection.

For example, an initial spin-up state can only mix with spin-down states
with orbital projection $l+1$, whereas an initial spin-down state can
only mix with spin-up states with orbital projection $l-1$.

The radial quantum number is not protected by this symmetry.  The
spin-fringe perturbation is linear in the transverse coordinate and
therefore changes the oscillator shell by one quantum at each insertion.
At first order this gives only the two radial sidebands displayed above.
At higher orders, repeated insertions can populate additional radial
levels inside the same fixed-$J_z$ sector.  The radial distribution is
therefore built as a perturbative ladder, while the orbital projection
remains restricted to the two values allowed by conservation of $J_z$.

In the perturbative regime, the $p$-th order contribution is suppressed
by the $p$-th power of the effective fringe strength.  If the accumulated
spin-flip probability becomes comparable to unity, the first-order
description is no longer sufficient, and one must solve the
$z$-ordered evolution in the corresponding fixed-$J_z$ subspace.

\begin{figure*}[t]
    \centering
    \includegraphics[width=1\linewidth]{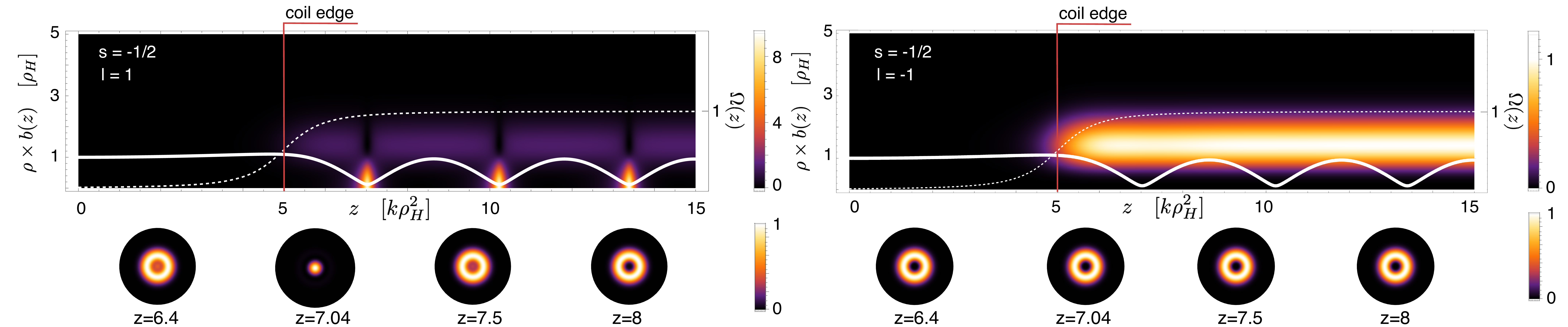}
 \caption{
Top row: normalized scaling parameter \(b(z)/b_0\) for the magnetic coil
edge, shown by the white solid line, and normalized magnetic-field profile
\(\Omega(z)\), shown by the white dashed line.  The solenoid entrance is
located at \(z_0=5\), and \(b(z)\) is calculated from
Eq.~\eqref{eq:ermakov_pinney} with \(b(0)=3\), \(b'(0)=0\).  The heat maps
show the normalized radial distribution of the first-order correction
\(\bigl|\varphi^{(1)}(b(z)\rho,0,z)\bigr|^2\), plotted as a function of
the comoving radius and the normalized propagation coordinate.  Bottom
row: transverse probability distributions of
\(\bigl|\varphi^{(1)}(b(z)\rho,0,z)\bigr|^2\) for representative values of
\(z\).  The distributions are normalized to the common prefactor
\(\bigl|s_q/(2k\rho_H b(z))\bigr|^2\).  The left column corresponds to the
initial state \(n=0\), \(l=1\), \(s=-1/2\), for which
Eq.~\eqref{eq:negspl} gives two allowed sidebands,
\(|0,0,\uparrow\rangle\) and \(|1,0,\uparrow\rangle\), leading to mode
beating.  The right column corresponds to the initial state \(n=0\),
\(l=-1\), \(s=-1/2\), for which Eq.~\eqref{eq:negsl} gives
only the single sideband \(|0,-2,\uparrow\rangle\).  In all panels,
\(\rho\) and \(z\) are measured in units of \(\rho_H\) and
\(k\rho_H^2\), respectively.}   
    \label{fig:Fig1}
\end{figure*}

To illustrate two distinct z-``evolution'' regimes of the correction
$\varphi^{(1)}(b(z)\rho,0;z)$, we consider the solenoid entrance with
the normalized magnetic field profile
\begin{equation}
\Omega(z)=\frac{1}{2}+\frac{1}{2}\frac{z-z_0}{\sqrt{1+(z-z_0)^2}}.
\label{eq:coilF}
\end{equation}
Here $z_0$ is the position of the solenoid entrance.  For $z_0=5$,
the resulting profile $\Omega(z)$ is shown in Fig.~\ref{fig:Fig1} by
the white dashed line.  We then solve the Ermakov-Pinney equation
\eqref{eq:ermakov_pinney} with the profile \eqref{eq:coilF} and initial
conditions $b(0)=b_0=3$, $b'(0)=b'_0=0$.  The corresponding scaling
parameter $b(z)$ is shown in Fig.~\ref{fig:Fig1} by the white solid
line.

Next, we consider a pure incoming state with well-defined spin projection
and orbital angular momentum before entering the coil.  In this case the
zeroth-order contribution $\varphi^{(0)}$ and the first-order correction
$\varphi^{(1)}$ are orthogonal in spin space, since the correction has
the opposite spin projection.  Therefore the probability density is
\begin{equation}
|\varphi|^2=|\varphi^{(0)}|^2+|\varphi^{(1)}|^2.
\end{equation}
As discussed in Refs.\cite{VanEnk2021,Greenshields,PhysRevA.108.012219}, $\varphi^{(0)}$ is a breathing mode whose transverse
scale and phase are governed by the Ermakov mapping.  In contrast,
$\varphi^{(1)}$ has a more complicated evolution along $z$, because
it is built from the spin-flipped modes allowed by the selection rules.

For the incoming state with $s=-1/2$, $l=1$, and $n=0$, the
correction consists of two modes,
\begin{equation}
|1,0,\uparrow\rangle,
\qquad
|0,0,\uparrow\rangle,
\end{equation}
as follows from Eq.~\eqref{eq:negspl}.  These two modes interfere, and
their relative phase changes during propagation through the edge.  To
illustrate this behavior, in the left part of Fig.~\ref{fig:Fig1} we plot
the radial probability density
$|\varphi^{(1)}(b(z)\rho,0,z)|^2$ and overlay it with the scaling
parameter $b(z)$.  One can see that minima of $b(z)$ roughly
correspond to a constructive combination close to
\begin{equation}
|1,0,\uparrow\rangle+|0,0,\uparrow\rangle,
\end{equation}
whereas away from these minima the correction is closer to the destructive
combination
\begin{equation}
|1,0,\uparrow\rangle-|0,0,\uparrow\rangle .
\end{equation}
Thus the oscillation of the Ermakov scale induces a beating between the
two allowed radial sidebands.

The situation is different for the incoming state with $s=-1/2$,
$l=-1$, and $n=0$.  In this case the correction contains only one
mode,
\begin{equation}
|0,-2,\uparrow\rangle,
\end{equation}
as follows from Eq.~\eqref{eq:negsl}.  Therefore no beating between
radial modes occurs.  This single-mode regime is shown in the right part
of Fig.~\ref{fig:Fig1}.

We note that this asymmetry is a property of the $n=0$ family (a family of incoming modes with zero radial excitation).  When
the signs of the spin projection and the orbital angular momentum are
opposite, two radial sidebands survive and the beating is present.  When
their signs are the same, one of the sidebands is eliminated by the lower
bound on the radial quantum number, and the correction contains only one
mode.

Finally, for a spin-unpolarized beam both spin sectors contribute, and
the solenoidal edge generates spin-orbital correlations in each allowed
sector.  The full evolution remains unitary, but if the orbital sidebands
are not resolved, the reduced spin state becomes effectively mixed.  This
effect can be enhanced after passage through several solenoid edges, where
additional radial sidebands are generated.

The mechanism discussed above is especially relevant for twisted electrons propagating through realistic accelerator optical lattices.  Such lattices contain sequences of solenoids, matching sections, and magnetic-field edges, so the transverse Pauli coupling can act repeatedly along the beam line.  If the lattice is not designed with the spin-orbital sidebands in mind, an initially well-defined twisted state can acquire coherent admixtures of spin-flipped orbital modes.  The full evolution remains unitary, but the purity of a selected spin or orbital subsystem can be degraded once the unresolved sidebands are traced out or filtered by apertures and diagnostics.  Conversely, this sensitivity can be turned into a constructive tool.  By engineering the positions, lengths, and strengths of magnetic edges, one can phase match the successive spin-fringe transitions and drive Rabi-type oscillations between selected spin-orbital modes.  Such a regime would provide a controlled mechanism for spin-OAM conversion and, combined with mode selection, could be used to generate spin-polarized twisted-electron states.

\section{Conclusion}

We have developed a paraxial theory of twisted-electron spin dynamics in
the fringe region of an axially symmetric magnetic coil. Starting from the
Foldy-Wouthuysen form of the Dirac equation, we derived the effective
spinor propagation equation in which the longitudinal field profile
generates both the usual diagonal orbital and Zeeman rotations and a
transverse Pauli coupling proportional to the field gradient. The scalar
part of the dynamics was treated exactly by the Ermakov mapping, while the
spin-fringe term was incorporated perturbatively. This yielded an explicit
first-order correction for arbitrary longitudinal solenoidal profiles
within the paraxial and perturbative regime.

The structure of the result is strongly constrained by axial symmetry. The
transverse Pauli interaction preserves the total axial angular momentum, so
a spin flip is necessarily accompanied by a compensating change of orbital
angular momentum. At first order, the linear coordinate form of the
perturbation restricts the dynamics to at most two neighboring radial
sidebands. The corresponding amplitudes are controlled by the
fringe-gradient strength dressed by the Ermakov scaling factor, whereas
their phases contain both the intrinsic Ermakov accumulation and the
diagonal spin-orbital rotation. This makes the final expressions
transparent and directly applicable to realistic magnetic-coil profiles.

Our analysis also clarifies the qualitative behavior of the generated
spin-flipped component. For $n=0$, depending on the incoming mode the correction may
contain either two interfering radial sidebands, leading to beating, or a
single allowed sideband. For spin-unpolarized or partially resolved beams,
the generated spin-orbital correlations imply that unresolved orbital
sidebands can reduce the purity of a selected spin or orbital subsystem
even though the full evolution remains unitary. This point is especially
relevant for twisted electrons propagating through realistic accelerator
optical lattices, where several magnetic edges may act successively.

The present framework shows that the fringe field provides a natural mechanism for coherent spin-orbit mode conversion. In an
undesigned lattice repeated interactions with the fringes can lead to unwanted mode mixing, whereas in an engineered sequence of magnetic-field edges it can, in principle, be
used constructively to phase match successive transitions and drive
Rabi-type oscillations between selected spin-orbital channels. This opens
a route toward controlled spin manipulation and toward the generation of
spin-polarized twisted-electron states in structured magnetic systems.

Although the present treatment was formulated for the axially symmetric
case, the underlying strategy is more general. The combination of exact
scalar evolution with perturbative treatment of the transverse Pauli
coupling can be extended to nonaxisymmetric fringe fields as well. In that
broader setting, the simple axial selection rules derived here will be
modified or lifted, but the same framework should allow one to analyze more
complex edge geometries containing higher-order multipole components. This
makes the present work a natural starting point for a systematic theory of
spin effects in realistic magnetic-lens and accelerator lattices, where
solenoidal, quadrupolar, skew, and higher multipolar fringe fields may all
contribute to twisted-electron mode conversion.

\bibliographystyle{apsrev4-2} 
\bibliography{ref}  

@article{Ivanov2011,
	author = {I. P. Ivanov},
	journal = {Physical Review D},
	pages = {093001},
	title = {Colliding particles carrying nonzero orbital angular momentum},
	volume = {83},
	year = {2011}}

@article{Ivanov2022,
	author = {I. P. Ivanov},
	journal = {Progress in Particle and Nuclear Physics},
	pages = {103987},
	title = {Promises and challenges of high-energy vortex states collisions},
	volume = {127},
	year = {2022}}

@article{Liu2025,
	author = {X. Liu and Q. Meng and Z. Yang and W. Ma and L. Lu and P. Zhang and L. Zou},
	journal = {Journal of Applied Physics},
	pages = {124303},
	title = {Design of a compact and integrated vortex field-emission electron source},
	volume = {138},
	year = {2025}}

@article{Karlovets2023,
	author = {D. V. Karlovets and S. S. Baturin and G. Geloni and G. K. Sizykh and V. G. Serbo},
	journal = {European Physical Journal C},
	pages = {372},
	title = {Shifting physics of vortex particles to higher energies via quantum entanglement},
	volume = {83},
	year = {2023}}

@article{Chaikovskaia2024,
	author = {A. D. Chaikovskaia and D. V. Karlovets and V. G. Serbo},
	journal = {Physical Review A},
	pages = {012222},
	title = {Vavilov-cherenkov emission with twisted electrons: a study of the final entangled state},
	volume = {109},
	year = {2024}}

@article{Pavlov2025,
	author = {I. Pavlov and A. Chaikovskaia and D. Karlovets},
	journal = {Physical Review C},
	pages = {024619},
	title = {Angular momentum effects in neutron decay},
	volume = {111},
	year = {2025}}

@article{Sheremet2025,
	author = {N. Sheremet and A. Chaikovskaia and D. Grosman and D. Karlovets},
	journal = {Physical Review A},
	pages = {052810},
	title = {Influence of the vortex-electron spatial distribution on atomic scattering},
	volume = {111},
	year = {2025}}

@article{Dyatlov2025,
	author = {A. S. Dyatlov and M. A. Nozdrin and A. N. Sergeev and N. E. Sheremet and S. S. Stafeev and D. V. Karlovets},
	journal = {Applied Optics},
	pages = {10776},
	title = {Generation of deep ultraviolet vortices via amplitude and phase spiral zone plates},
	volume = {64},
	year = {2025}}

@misc{RFPhotoinjector2025,
	archiveprefix = {arXiv},
	author = {A. S. Dyatlov and A. V. Afanasyev and V. V. Kobets and A. E. Levichev and M. V. Maksimov and D. A. Nikiforov and M. A. Nozdrin and K. Popov and K. A. Sibiryakova and K. E. Yunenko and D. V. Karlovets},
	eprint = {2509.00732},
	title = {Classical and quantum beam dynamics simulation of the {RF} photoinjector test bench},
	year = {2025}}

@article{Greenshields,
	author = {Greenshields, Colin R. and Stamps, Robert L. and Franke-Arnold, Sonja and Barnett, Stephen M.},
	doi = {10.1103/PhysRevLett.113.240404},
	issue = {24},
	journal = {Phys. Rev. Lett.},
	month = {Dec},
	numpages = {5},
	pages = {240404},
	publisher = {American Physical Society},
	title = {Is the Angular Momentum of an Electron Conserved in a Uniform Magnetic Field?},
	url = {https://link.aps.org/doi/10.1103/PhysRevLett.113.240404},
	volume = {113},
	year = {2014}}

@misc{EpovArxiv,
	archiveprefix = {arXiv},
	author = {M. S. Epov and I. E. Shenderovich and S. S. Baturin},
	eprint = {2602.07858},
	primaryclass = {quant-ph},
	title = {Geometry-Enabled Radiation from Structured Paraxial Electrons},
	url = {https://arxiv.org/abs/2602.07858},
	year = {2026}}

@article{FW,
	author = {Foldy, Leslie L. and Wouthuysen, Siegfried A.},
	doi = {10.1103/PhysRev.78.29},
	issue = {1},
	journal = {Phys. Rev.},
	month = {Apr},
	numpages = {0},
	pages = {29--36},
	publisher = {American Physical Society},
	title = {{On the Dirac theory of spin 1/2 particles and its non-relativistic limit}},
	url = {https://link.aps.org/doi/10.1103/PhysRev.78.29},
	volume = {78},
	year = {1950}}

@article{SilenkoFW,
	author = {Silenko, Alexander J.},
	doi = {10.1063/1.1579991},
	issn = {0022-2488},
	journal = {Journal of Mathematical Physics},
	month = {06},
	number = {7},
	pages = {2952-2966},
	title = {{Foldy--Wouthuysen transformation for relativistic particles in external fields}},
	volume = {44},
	year = {2003}}

@article{SilenkoG,
	author = {Silenko, A. J.},
	doi = {10.1142/S0217732322500973},
	journal = {Modern Physics Letters A},
	number = {16},
	pages = {2250097},
	title = {Exact quantum-mechanical equations for particle beams},
	volume = {37},
	year = {2022}}

@book{itzykson1985quantum,
	author = {Itzykson, C. and Zuber, J.B.},
	isbn = {9780070663534},
	lccn = {78040977},
	publisher = {McGraw-Hill},
	series = {International series in pure and applied physics},
	title = {Quantum Field Theory},
	url = {https://books.google.com.cy/books?id=46m8QgAACAAJ},
	year = {1985}}

@book{LandauQM,
	address = {Oxford},
	author = {Landau, L. D. and Lifshitz, E. M.},
	edition = {3},
	publisher = {Pergamon Press},
	series = {Course of Theoretical Physics},
	title = {Quantum Mechanics: Non-Relativistic Theory},
	volume = {3},
	year = {1991}}

@book{Sakurai2017,
	address = {Cambridge},
	author = {Sakurai, J. J. and Napolitano, Jim},
	edition = {2},
	publisher = {Cambridge University Press},
	title = {Modern Quantum Mechanics},
	year = {2017}}

@article{VanEnk2021,
	author = {Melkani, Abhijeet and van Enk, S. J.},
	doi = {10.1103/PhysRevResearch.3.033060},
	issue = {3},
	journal = {Phys. Rev. Res.},
	month = {Jul},
	numpages = {8},
	pages = {033060},
	publisher = {American Physical Society},
	title = {Electron vortex beams in nonuniform magnetic fields},
	url = {https://link.aps.org/doi/10.1103/PhysRevResearch.3.033060},
	volume = {3},
	year = {2021}}

@article{Filina2026,
	author = {Filina, N. V. and Baturin, S. S.},
	doi = {10.1103/y1d1-dzqh},
	issue = {2},
	journal = {Phys. Rev. A},
	month = {Feb},
	numpages = {6},
	pages = {L021302},
	publisher = {American Physical Society},
	title = {Universal analytic solution for the quantum transport of structured matter waves in magnetic optics},
	url = {https://link.aps.org/doi/10.1103/y1d1-dzqh},
	volume = {113},
	year = {2026}}

@article{Filina2024,
	author = {Filina, N. V. and Baturin, S. S.},
	doi = {10.1103/PhysRevA.110.022204},
	issue = {2},
	journal = {Phys. Rev. A},
	month = {Aug},
	numpages = {8},
	pages = {022204},
	publisher = {American Physical Society},
	title = {Twisted charged particles in the uniform magnetic field with broken symmetry},
	url = {https://link.aps.org/doi/10.1103/PhysRevA.110.022204},
	volume = {110},
	year = {2024}}

@article{Bliokh2007,
	author = {Bliokh, Konstantin Yu. and Bliokh, Yury P. and Savel'ev, Sergey and Nori, Franco},
	doi = {10.1103/PhysRevLett.99.190404},
	journal = {Physical Review Letters},
	pages = {190404},
	title = {Semiclassical Dynamics of Electron Wave Packet States with Phase Vortices},
	volume = {99},
	year = {2007}}

@article{Uchida2010,
	author = {Uchida, Masaya and Tonomura, Akira},
	doi = {10.1038/nature08904},
	journal = {Nature},
	number = {7289},
	pages = {737--739},
	title = {Generation of Electron Beams Carrying Orbital Angular Momentum},
	volume = {464},
	year = {2010}}

@article{Verbeeck2010,
	author = {Verbeeck, Jo and Tian, Hui and Schattschneider, Peter},
	doi = {10.1038/nature09366},
	journal = {Nature},
	number = {7313},
	pages = {301--304},
	title = {Production and Application of Electron Vortex Beams},
	volume = {467},
	year = {2010}}

@article{McMorran2011,
	author = {McMorran, Benjamin J. and Agrawal, Amit and Anderson, Ian M. and Herzing, Andrew A. and Lezec, Henri J. and McClelland, Jabez J. and Unguris, John},
	doi = {10.1126/science.1198804},
	journal = {Science},
	number = {6014},
	pages = {192--195},
	title = {Electron Vortex Beams with High Quanta of Orbital Angular Momentum},
	volume = {331},
	year = {2011}}

@article{Lloyd2017,
	author = {Lloyd, S. M. and Babiker, M. and Thirunavukkarasu, G. and Yuan, J.},
	doi = {10.1103/RevModPhys.89.035004},
	journal = {Reviews of Modern Physics},
	pages = {035004},
	title = {Electron Vortices: Beams with Orbital Angular Momentum},
	volume = {89},
	year = {2017}}

@article{Bliokh2017,
	author = {Bliokh, Konstantin Y. and Ivanov, Igor P. and Guzzinati, Giulio and Clark, Laura and Van Boxem, Ruben and B{\'e}ch{\'e}, Armand and Juchtmans, Roeland and Alonso, Miguel A. and Schattschneider, Peter and Nori, Franco and Verbeeck, Jo},
	doi = {10.1016/j.physrep.2017.05.006},
	journal = {Physics Reports},
	pages = {1--70},
	title = {Theory and Applications of Free-Electron Vortex States},
	volume = {690},
	year = {2017}}

@article{Silenko2019,
	author = {Silenko, Alexander J. and Zhang, Pengming and Zou, Liping},
	doi = {10.1103/PhysRevLett.122.063201},
	journal = {Physical Review Letters},
	pages = {063201},
	title = {Electric Quadrupole Moment and the Tensor Magnetic Polarizability of Twisted Electrons and a Potential for Their Measurements},
	volume = {122},
	year = {2019}}

@article{PhysRevA.108.012219,
	author = {Filina, N. V. and Baturin, S. S.},
	doi = {10.1103/PhysRevA.108.012219},
	issue = {1},
	journal = {Phys. Rev. A},
	month = {Jul},
	numpages = {13},
	pages = {012219},
	publisher = {American Physical Society},
	title = {Unitary equivalence of twisted quantum states},
	url = {https://link.aps.org/doi/10.1103/PhysRevA.108.012219},
	volume = {108},
	year = {2023}}

@article{PhysRevA.103.L010201,
	author = {Zou, Liping and Zhang, Pengming and Silenko, Alexander J.},
	doi = {10.1103/PhysRevA.103.L010201},
	issue = {1},
	journal = {Phys. Rev. A},
	month = {Jan},
	numpages = {7},
	pages = {L010201},
	publisher = {American Physical Society},
	title = {General quantum-mechanical solution for twisted electrons in a uniform magnetic field},
	url = {https://link.aps.org/doi/10.1103/PhysRevA.103.L010201},
	volume = {103},
	year = {2021}}

@article{Aldaya_2011,
	author = {Aldaya, V and Coss{\'\i}o, F and Guerrero, J and L{\'o}pez-Ruiz, F F},
	doi = {10.1088/1751-8113/44/6/065302},
	journal = {Journal of Physics A: Mathematical and Theoretical},
	month = {jan},
	number = {6},
	pages = {065302},
	title = {The quantum Arnold transformation},
	url = {https://doi.org/10.1088/1751-8113/44/6/065302},
	volume = {44},
	year = {2011}}
\end{document}